\documentclass[prb,amsmath,a4paper,twocolumn]{revtex4}

\usepackage{epsfig}
\usepackage{graphicx}  

\usepackage{color}

\newcommand{\I}{\mathrm{i}}

\newcommand{\be}{\begin{equation}}
\newcommand{\ee}{\end{equation}}
\newcommand{\bea}{\begin{eqnarray}}
\newcommand{\eea}{\end{eqnarray}}
\newcommand{\ba}{\begin{array}}
\newcommand{\ea}{\end{array}}

\newcommand{\ELF}{{\rm ELF} }

\begin{document}

\title{Electron localization function for two-dimensional systems}

\author{E. R{\"{a}}s{\"{a}}nen}
\email[Electronic address:\;]{esa@physik.fu-berlin.de}
\affiliation{
Institut f{\"u}r Theoretische Physik,
Freie Universit{\"a}t Berlin, Arnimallee 14, D-14195 Berlin, Germany}
\affiliation{European Theoretical Spectroscopy Facility (ETSF)}

\author{A. Castro}
\email[Electronic address:\;]{alberto@physik.fu-berlin.de}
\affiliation{
Institut f{\"u}r Theoretische Physik,
Freie Universit{\"a}t Berlin, Arnimallee 14, D-14195 Berlin, Germany}
\affiliation{European Theoretical Spectroscopy Facility (ETSF)}

\author{E. K. U. Gross} 
\affiliation{
Institut f{\"u}r Theoretische Physik,
Freie Universit{\"a}t Berlin, Arnimallee 14, D-14195 Berlin, Germany}
\affiliation{European Theoretical Spectroscopy Facility (ETSF)}


\begin{abstract}
  The concept of the electron localization function (ELF) is 
  extended to two-dimensional (2D) electron systems. We show that the
  topological properties of the ELF in 2D are 
  considerably simpler than in molecules
  studied previously. We compute the ELF and demonstrate its usefulness
  for various physical 2D systems, focusing on semiconductor quantum 
  dots that effectively correspond to a confined 2D electron gas.
  The ELF visualizes the shell structure of harmonic quantum
  dots and provides insight into electron bonding 
  in quantum-dot molecules. In external magnetic fields
  the ELF is found to be a useful measure of vorticity 
  when analyzing the properties of quantum-Hall droplets.
  We show that the current-dependent term in the ELF 
  expression is important in magnetic
  fields.
\end{abstract}


\maketitle

\section{Introduction}
\label{section:introduction}

The study of electron localization is linked to the pursuit of a
rigorous systematization of the elusive concept of a chemical
bond,\cite{pauling-1948, lewis-1966} ubiquitous in quantum chemistry
ever since the first attempts of its description.\cite{lewis-1916} 
The main point in the concept of electron localization is not where 
the electrons {\em are} (or in quantum terms 
where the electrons are likely 
to be), which can be monitored from the electron density, 
but where the electrons {\em are localized}. 

The concept of localization (and delocalization) arises 
from Pauli's exclusion principle: two electrons with
the same spin cannot occupy the same spatial position. The consequence
is the appearance of the {\em Fermi hole},~\cite{lowdin-1959} a
function which determines how the probability of finding an electron
at some point diminishes because of the presence of another like-spin
electron in its vicinity. In other words,
the Fermi hole is that part of the exchange
and correlation hole that stems solely from the fermionic nature of
the electron and not because of the Coulomb repulsion. 
Any electron at a given point carries its associated Fermi hole as a
counterpart. 
The localization or delocalization of the
electron is equivalent to the localization or delocalization of its
Fermi hole.~\cite{bader-1996, bader-1996-II} Naturally, a
localization descriptor should be based on this function or,
correspondingly, on the like-spin pair probability function -- this will be
clarified below. Indeed, the so-called electron localization function
(ELF)~\cite{becke-1990, silvi-1994, savin-1997, silvi-2005} is nothing
else than an appropriate renormalization of the Fermi hole
curvature.\cite{dobson-1991, dobson-1993}

The ELF is large where electrons {\em pair}. Our intuition in chemistry
tells us that electrons should pair forming localized
groups in each atomic shell of the inert cores, in the chemical
bonds, and in the non-bonding or {\em lone} pairs. More precisely, the
topological properties of a good localization function should
partition the space for each group of electrons -- also for
delocalized groups of electrons such as $\pi$ systems in conjugated
molecules, for which the ELF values should be low. Our understanding
of chemistry is founded on the electron {\em pair}, a consequence of
Pauli's exclusion principle, and more particularly of the localization
of one electron of each spin in a region of space. Accordingly, it is
not surprising that the main purpose of the analysis of the ELF has
been to help our intuition of the fundamental chemistry concepts
of pairs and bonds. 


Due to this focus on the elucidation of the chemical bond, all studies
of electron localization functions performed to date have referred to
three-dimensional (3D) systems. To our knowledge, no
attempts have been made to map the electron localization in
two dimensions (2D). However, significant advances in semiconductor technology 
have enabled the production and manipulation of low-dimensional
nanoscale structures. Common examples of these systems
are quantum dots (QDs),~\cite{qd-review} 
often also called artificial atoms. The electrons in QDs are confined
in the interface of a semiconductor heterostructure,
e.g., GaAs/AlGaAs, so that the transverse dimensions
controlled by a lateral confinement are considerably larger
than the thickness of the QD. In most cases a 2D model
describes the movement of the electrons with a 
reasonable accuracy. On the other hand, the approximation 
for the in-plane confinement depends on the QD shape that 
can be, e.g., circular, rectangular, ring-like (quantum ring), 
or consist of several potential minima (quantum-dot molecule).
Due to the tunable shape, size, and number of electrons, QDs
have emerging applications in fields of quantum computation,
and from a theoretical point of view they are an ideal
playground to study many-electron phenomena and test 
computational methods. Hence, we expect the ELF 
to be of interest for various QD systems, since it carries valuable
information about the electronic structure regardless of the
dimensionality. 

This article is organized as follows. In Sec.~\ref{section:elf}
we present a detailed derivation of the ELF in 2D, 
focusing on the definition of the Fermi-hole curvature 
which is the core of the ELF, and show how the relevant 
expressions hold compared with the 3D case. 
We also discuss the topological
properties of the ELF in 2D. 
In Sec.~\ref{sec:model} we briefly present
the QD model and our computational scheme based on 
spin-density-functional theory. 
In Sec.~\ref{sec:examples} 
we provide four examples where the ELF is a useful tool
when analyzing the electronic properties of QDs. They include
(i) visualization of the shell structure of a 
parabolic (harmonic) QD, (ii) showing the bond-like 
interdot electron couplings in QD molecules, (iii) 
introducing the ELF as a measure of vorticity in
quantum-Hall droplets, and (iv) visualization of vortex
localization in high magnetic fields.
We also show that the current-dependent term in the 
ELF (or in the Fermi-hole curvature), which has
been neglected in most studies until now, 
is relevant in order to obtain meaningful results
for QDs in magnetic fields.
Concluding remarks are given in Sec.~\ref{sec:summary}.

\section{Electron localization function}
\label{section:elf}


\subsection{Derivation of the ELF in two dimensions}

In the following we go in detail through the mathematical
derivation of the ELF. We focus on the differences in the expressions
in 3D and 2D, in particular between the spherical and circular
averages and kinetic-energy densities. 


We consider the many-electron wave function 
$\Psi({\bf r}_1\sigma_1,\dots,{\bf r}_N\sigma_N)$,
where $N$ is the number of electrons, and ${\bf r}$ and $\sigma$ are the
electronic position and spin coordinates, respectively.
We point out that $\Psi$ can be either an eigenstate
of a static Hamiltonian, or an evolving time-dependent
state. Nevertheless, we omit the time coordinate in the 
notation. The necessary ingredients in the derivation of the ELF 
are the first-order reduced density
matrix~\cite{lowdin-1955,mcweeny-1960,mcweeny-1989} 
\begin{eqnarray}
\lefteqn{\Gamma^{(1)}_{\sigma_1\sigma_1'}({\bf r}_1 \vert {\bf r}'_1) = N\sum_{\sigma_2,\dots,\sigma_N}\int\, {\rm d} {\bf r}_2\dots{\rm d} {\bf r}_N \times} \nonumber \\
& & \Psi^{*}({\bf r}_1\sigma_1,\dots,{\bf r}_N\sigma_N)\Psi({\bf r}'_1\sigma'_1,\dots,{\bf r}_N\sigma_N)\,,
\end{eqnarray}
and the diagonal of the second-order reduced density 
matrix\footnote{Note that we follow the convention of McWeeny,\cite{mcweeny-1960} who uses a $N(N-1)$ factor for the second order density matrix, rather than the $N(N-1)/2$ factor used by L{\"{o}}wdin.\cite{lowdin-1955}}
\begin{eqnarray}
\lefteqn{\gamma^{(2)}_{\sigma_1\sigma_2}({\bf r}_1,{\bf r}_2) = } \nonumber \\
& & N(N-1)\sum_{\sigma_3,\dots,\sigma_N} \int {\rm d} {\bf r}_3\dots{\rm d} {\bf r}_N \times \nonumber \\
& & \vert\Psi({\bf r}_1\sigma_1,{\bf r}_2\sigma_2,\dots,{\bf r}_N\sigma_N)\vert^2.
\label{eq:gamma2}
\end{eqnarray}
The diagonal of $\Gamma^{(1)}$ is commonly known as the spin-density, 
\begin{equation}
n_{\sigma}({\bf r}) = \Gamma^{(1)}_{\sigma\sigma}({\bf r} \vert {\bf r}),
\label{spindensity}
\end{equation}
and the total density, $n({\bf r}) = \sum_{\sigma}n_{\sigma}({\bf r})$, sums over
the spin variable. It has an easy interpretation: $n({\bf r}){\rm
  d}{\bf r}$ is the probability (normalized to $N$) of finding an electron 
in a small volume ${\rm
  d}{\bf r}$ at position ${\bf r}$. The {\em pair probability density}
 is defined analogously by 
\begin{equation}
n^{(2)}({\bf r}_1, {\bf r}_2) = \sum_{\sigma_1,\sigma_2}
\gamma^{(2)}_{\sigma_1\sigma_2}({\bf r}_1,{\bf r}_2)
\end{equation}
and can be interpreted in the following way: $n^{(2)}({\bf r}_1, {\bf r}_2) {\rm
  d}{\bf r}_1 {\rm d}{\bf r}_2 $ is the probability of finding one electron at
${\bf r}_1$ and, {\em simultaneously}, another electron at ${\bf
  r}_2$.

Were the electrons fully independent, this latter probability would
just be the product of the electron densities: $n^{(2)}({\bf r}_1,
{\bf r}_2) = n({\bf r}_1)n({\bf r}_2)$. However, their exchange
(fermionic character) and Coulomb interaction reduce the pair
probability by a value known as the exchange and correlation hole:
\begin{equation}
n^{(2)}({\bf r}_1,
{\bf r}_2) = n({\bf r}_1)n({\bf r}_2) + h({\bf r}_1, {\bf r}_2)\,.
\end{equation}
The Fermi hole is the part of $h({\bf r}_1, {\bf r}_2)$ which is
entirely due to the antisymmetric character of the wave function,
regardless of the Coulomb interaction. Accordingly, it only appears
for like-spin electrons. It is the dominant part
at short distances (${\bf r}_2 \to {\bf r}_1$).

The key function
to study the electron localization is the like-spin conditional pair
probability function defined as
\begin{equation}
\label{eq:condpair}
P_{\sigma}({\bf r}_1,{\bf r}_2) \equiv
\frac{\gamma^{(2)}_{\sigma\sigma}({\bf r}_1,{\bf r}_2)}{n_{\sigma}({\bf r}_1)}\,.
\end{equation}
The physical meaning of this function is the following. It gives the
probability of finding one $\sigma$-spin electron at ${\bf r}_2$
knowing with certainty that another $\sigma$-spin electron is at ${\bf
r}_1$. The question is now how probable it is to find one like-spin
electron in the vicinity of the first reference one.
For this purpose it is useful to define a conditional
pair probability function at a distance $s$ via a spherical (in 3D) or
circular (in 2D) average. Here we deal with the latter case, 
\begin{equation}
\label{eq:condpairdist}
p_{\sigma}({\bf r}_1, s) = \frac{1}{2\pi}\int_0^{2\pi}\!\! {\rm d}\theta\;
P_{\sigma}({\bf r}_1,{\bf r}_1 + s\hat{\bf u}_\theta)\,,
\end{equation}
where $\hat{\bf u}_\theta = \cos\theta\hat{\bf x}_1 + \sin\theta\hat{\bf x}_2$.
Since we are particularly interested in the behavior of 
$p_\sigma({\bf  r},s)$ at small $s$, we need to perform Taylor expansions,
\begin{eqnarray}
\nonumber
\lefteqn{\Psi({\bf r}_1\sigma, ({\bf r}_1 + s\hat{\bf u}_\theta)\sigma,\dots,{\bf r}_N\sigma_N) =} 
\\
& & s\, ( \hat{\bf u}_\theta \cdot \nabla_{2})
\Psi({\bf r}_1\sigma,{\bf r}_2\sigma,\dots,{\bf r}_N\sigma_N)_{{\bf r}_2={\bf r}_1}
+ \mathcal{O}(s^2)\,,
\end{eqnarray}
where $\nabla_2$ is the gradient with respect to the second electron
variable in $\Psi$. The $s^0$ term is absent due to Pauli's exclusion
principle. We now take the square,
\begin{eqnarray}
\nonumber
\lefteqn{\vert \Psi({\bf r_1}\sigma, ({\bf r}_1 + s\hat{\bf u}_\theta)\sigma,\dots,{\bf r}_N\sigma_N)\vert^2 = } \\
& & s^2 \, \left|(\hat{\bf u}_\theta \cdot \nabla_2)\Psi({\bf r}_1\sigma,{\bf r}_2\sigma,\dots,{\bf r}_N\sigma_N)_{{\bf r}_2={\bf r}_1}\right|^2 + \mathcal{O}(s^3) = \nonumber \\
& & s^2 \sum_{i,j=1}^{2} c^*_i c_j u_i(\theta)u_j(\theta) + \mathcal{O}(s^3)\,,
\end{eqnarray}
where
\begin{equation}
c_i = (\hat{\bf x}_i \cdot \nabla_2) \Psi({\bf r}_1\sigma,{\bf r}_2\sigma,\dots,{\bf r}_N\sigma_N)_{{\bf r}_2={\bf r}_1}
\end{equation}
with $i=1,2$. The circular average is
\begin{eqnarray}
\lefteqn{\frac{1}{2\pi}\int_0^{2\pi}\!\! {\rm d}\theta\;
\vert \Psi({\bf r_1}\sigma, ({\bf r}_1 + s\hat{\bf u}_\theta)\sigma,\dots,{\bf r}_N\sigma_N)\vert^2 = } \nonumber \\
& & s^2 \sum_{i,j=1}^{2} c^*_i c_j \frac{1}{2\pi}\int_0^{2\pi}\!\! {\rm d}\theta\; u_i(\theta)u_j(\theta) = \nonumber \\
& & \frac{1}{2}s^2\sum_{i=1}^{2} \vert c_i\vert^2 + \mathcal{O}(s^3) = \nonumber \\ 
& & \frac{1}{4}s^2 \nabla^2_2 \vert\Psi({\bf r}_1\sigma,{\bf r}_2\sigma,\dots,{\bf r}_N\sigma_N)\vert^2_{{\bf r}_2={\bf r}_1}
+\mathcal{O}(s^3)
\end{eqnarray}
This expression may then be used in combination with
Eqs.~(\ref{eq:gamma2}), (\ref{eq:condpair}), and (\ref{eq:condpairdist}) to
obtain
\begin{eqnarray}
\label{eq:condpairdist2}
p_\sigma({\bf r}_1, s) =
\frac{1}{2}s^2 C_{\sigma}({\bf r}_1) + \mathcal{O}(s^3)\,,
\\
\label{eq:csigma}
C_{\sigma}({\bf r}_1) = \frac{1}{2} \frac{\nabla^2_2 \gamma^{(2)}_{\sigma\sigma}
({\bf r}_1,{\bf r}_2)_{{\bf r}_2={\bf r}_1}}{n_{\sigma}({\bf r}_1)}.
\end{eqnarray}
These equations are similar to the ones obtained in the 3D case, in fact, 
they are the same {\em except}
for the factor $1/2$ in Eq.~(\ref{eq:condpairdist2}), which is
$1/3$ in 3D. The function $C_{\sigma}({\bf r})$ 
satisfies exactly
the same Eq.~(\ref{eq:csigma}) in 2D and in 3D.
This is the function that measures the {\em local}
like-spin pair probability, is an inverse measure of electron
localization, and is the function used to define the ELF.  We should
note also that $\nabla^2_2 \gamma^{(2)}_{\sigma\sigma} ({\bf
  r}_1,{\bf r}_2)_{{\bf r}_2={\bf r}_1} $ is the Fermi hole curvature
as defined by Dobson.~\cite{dobson-1991} Other definitions, equivalent
regarding the information that they contain, are
also possible.~\cite{stoll-1980}

We conclude that the expressions that define the ELF do not change
significantly from 3D to 2D.  The function $C_\sigma({\bf r})$ alone
does not reveal the localization explicitly. For that purpose one
needs to perform the following renormalization, which defines the ELF
as~\footnote{We point out that an alternative route to define a
localization function has been described by Kohout.~\cite{kohout} For
one-determinantal wave functions Kohout's {\em electron
localizability index} reduces to the ELF expression, with
no need of an {\em ad-hoc} renormalization.}
\begin{equation}
\ELF_{\sigma}({\bf r}) = 
\frac{1}{1 + 
\left\{C_{\sigma}({\bf r})/C_{\sigma}^{\rm HEG}[n_{\sigma}({\bf r})]   \right\}^2
}\,,
\label{eq:ELF}
\end{equation}
where $C_{\sigma}^{\rm HEG}[n_{\sigma}({\bf r})]$ is the value of the
$C_\sigma$ function for the homogeneous electron gas (HEG) of (constant)
spin densities $n_{\sigma}({\bf r})$. 
Analogously to the 3D case, the expression for 
$C_\sigma^{\rm HEG}$ in 2D is
the kinetic-energy density,
\begin{equation}
C_{\sigma}^{\rm HEG}[n_\sigma] = \tau_\sigma[n_\sigma] = 2\pi n_\sigma\,.
\end{equation}
In terms of the total density $n$ and the polarization
(often called also magnetization) $\zeta = (n_\uparrow-n_\downarrow)/n$,
the expression immediately yields 
\begin{equation}
C_{\sigma}^{\rm HEG}[n,\zeta] =\frac{\pi}{2}(1+\zeta^2)n
\end{equation}
for the total kinetic-energy density.

The above definition of $C_{\sigma}$ is completely general, but
requires the diagonal of the second-order reduced density matrix,
which is a fairly complex and often unmanageable object.  
In most cases, however, a further approximation is taken by
assuming that $\Psi$ is a single-determinantal wave function. 
In practice, this means that one uses the Hartree-Fock 
approximation, therefore neglecting the effect of correlations 
in the \ELF, or one uses density-functional theory (DFT)
(see below), and calculates the \ELF of the non-interacting 
auxiliary Kohn-Sham (KS) system. Both
possibilities are in principle unsatisfactory, but the experience 
has demonstrated that the results are usually unaffected
despite a notable mathematical simplification (see
Ref.~[\onlinecite{burnus-2004}] for the detailed steps).
If $\Psi$ is formed by the set of KS orbitals
$\lbrace\varphi_{i\uparrow}\rbrace_{i=1}^{N_\uparrow}$ and
$\lbrace\varphi_{i\downarrow}\rbrace_{i=1}^{N_\downarrow}$
($N_\uparrow+N_\downarrow = N$) within the spin-DFT (SDFT), 
one finds
\begin{equation}
\label{eq:csigma1det}
C_{\sigma}({\bf r}) = \tau_{\sigma}({\bf r})
-\frac{1}{4}\frac{(\nabla n_{\sigma}({\bf r}))^2}{n_{\sigma}({\bf r})}
-\frac{{\mathbf j}_{p,\sigma}^2({\bf r})}{n_{\sigma}({\bf r})}\,,
\end{equation}
where $n_{\sigma}({\bf r}) = \sum_{i=1}^{N_\sigma} \vert \varphi_{i\sigma}({\bf r})\vert^2$ is the spin-density [cf. Eq.~(\ref{spindensity})],
$\tau_{\sigma}$ is the kinetic-energy density given by
\begin{equation}
\tau_{\sigma}({\bf r}) = \sum_{i=1}^{N_\sigma} \vert \nabla\varphi_{i\sigma}({\bf r})\vert^2\,,
\label{tau_ks}
\end{equation}
and ${\mathbf j}_{p,\sigma}$ is the spin-resolved paramagnetic current density
\begin{equation}
{\mathbf j}_{p,\sigma}({\bf r}) = \frac{\I}{2}\sum_{i=1}^{N_\sigma} 
\left[\left(
\nabla\varphi^*_{i\sigma}({\bf r})\right)\varphi_{i\sigma}({\bf r})
- \varphi^*_{i\sigma}({\bf r}) \nabla\varphi_{i\sigma}({\bf r})
\right].
\label{current}
\end{equation}
Here we point out that most ELF calculations refer to 
closed-shell molecules (or finite systems in general) in their ground state 
and in the absence of magnetic fields or the spin-orbit coupling.
In those cases real wave functions can be assumed and
the current term in Eq.~(\ref{eq:csigma1det}) is absent.
In Secs.~\ref{subsection:mdd} and \ref{subsection:vortex} 
we note 
the importance of the current term, which is needed in the ELF
also during time-dependent processes.\cite{burnus-2005,castro-2007}

\subsection{Topological properties}
\label{subsection:topology}

The basis for the topological investigation of the ELF was
established by Silvi and Savin.~\cite{silvi-1994} The information
contained in the \ELF is extracted with the help of some basic
concepts borrowed from the theory of dynamical 
systems.\cite{abraham-1992,abraham-1994} 
The ELF is a scalar continuous
function bounded between 0 and 1, and one should look at its gradient
field, which in turn defines the set of attractors -- 
roughly speaking the maxima of the ELF. The
important concept is then that of a {\em basin} of each attractor -- the
set of points for which that attractor is the $\omega$-limit. 
In chemistry, each basin is then identified either as a 
core basin (it contains a nucleus), or as a valence basin. 
The latter ones are then classified
according to the number of core basins that they have frontier with:
if they have only one neighbor core, they are called monosynaptic; if
they have more than one (disynaptic, trisynaptic etc.), they are
{\em bonds}. The absence of real nuclei in the 2D models makes this
distinction not relevant. However, the division of space into
localization basins is still pertinent: the basins give us information about
the {\em groups} of electrons; exchange of electrons is more unlikely
between different basins than inside them. In 3D the attractors can
then be single points in the absence of special symmetries,
ring-shaped lines of attractors around a symmetry axis, or spheres
around isolated atoms. In 2D, obviously, the topology simplifies and
we can only find, besides point attractors, ring attractors for
cylindrically symmetrical problems.

\section{Quantum-dot model}
\label{sec:model}
We study the ELF in 2D QDs restricted to the {\em xy} plane.
The system is described by an effective-mass $N$-electron Hamiltonian
\begin{equation}
H = \sum^N_{i=1}\left[\frac{\left({\mathbf p}_i+
\frac{1}{c}{\bf A}_i\right)^2}{2m^*}
+V_{\rm ext}({\mathbf r}_i) + E_{Z,i} \right]
+ \sum_{i<j}\frac{1}{\epsilon^* |{\mathbf r}_i-{\mathbf r}_j|},
\label{hamiltonian}
\end{equation}
where ${\bf A}$ is the external vector potential 
(in symmetric gauge) of the homogeneous magnetic field 
${\bf B}=B_0{\hat z}$ perpendicular
to the plane, $V_{{\rm ext}}({\mathbf r})$ is the
external confining potential in the {\em xy} plane (see below),
and $E_Z=g^*\mu_B s_{z}B_0$ is the Zeeman energy.
We apply the conventional effective-mass 
approximation with the material parameters for GaAs:
the effective mass $m^*=0.067$, the dielectric constant 
$\epsilon^*=12.4$, and the gyromagnetic ratio $g^*=-0.44$.

We solve the ground-state problem associated with
the $N$-electron Hamiltonian~(\ref{hamiltonian})
by applying the SDFT in the collinear-spin representation.
The KS states, needed in Eqs.~(\ref{eq:csigma1det})-(\ref{current})
for computing the ELF as defined in Eq.~(\ref{eq:ELF}),
are solved from the KS equation
\begin{equation}
\left[\frac{({-i\nabla}_i+\frac{1}{c}{\mathbf A}({\mathbf r}) )^2}{2 m^*}
+V_{\rm KS}^{\sigma}({\mathbf r})\right]
\varphi_{i\sigma}(\mathbf{r})=\epsilon_{i\sigma}\varphi_{i\sigma}(\mathbf{r}),
\label{eq:kse}
\end{equation}
where the KS potential $V_{\rm KS}^{\sigma}(\mathbf{r})$ is a sum 
of the external confining potential, the Hartree potential, and the
exchange-correlation potential given by 
$V_{\rm xc}^{\sigma}({\mathbf r})=\delta{E_{\rm xc}}[n^{\uparrow},n^{\downarrow}]/\delta n^\sigma{({\mathbf r})}$.  
To approximate the exchange-correlation energy $E_{\rm xc}$
we use the local spin-density approximation (LSDA) with a parametrization
provided by Attaccalite and co-workers.~\cite{attaccalite} 
In QD systems the SDFT scheme together with the LSDA 
leads to good numerical accuracy in comparison with quantum 
Monte Carlo calculations, even in relatively high magnetic 
fields.~\cite{LDAtesting,spin-droplet}
We also point out that the local-vorticity approximation within the 
current-SDFT~\cite{vignale} does not lead to a considerable improvement 
over the SDFT results.~\cite{LDAtesting} In the numerical calculations 
we apply the real-space {\tt octopus} code.~\cite{octopus}

\section{Examples}
\label{sec:examples}

\subsection{Shell structure}
\label{subsection:shell}

First we examine the ELF in parabolic QDs at
zero magnetic field ($B=0$). Thus, we choose
the external confining potential in
Eq.~(\ref{hamiltonian}) as
$V_{{\rm ext}}(r)=\omega_0^2 r^2/2$, where
we set the confinement strength to
$\omega_0=0.42168$ a.u. $=5$ meV.
The system is a 2D harmonic oscillator
with single-electron eigenenergies
$\epsilon_{nl}=(2n+|l|+1)\omega_0$, where 
$n=0,1,2,\ldots$ is the radial and
$l=0,\pm 1,\pm 2,\ldots$ is the 
azimuthal quantum number.
The corresponding shell structure has been
experimentally well depicted in the addition energies 
of a double-barrier GaAs QD.~\cite{tarucha}
Further details in the measured addition
energies resulting from the electron-electron
interaction, such as Hund's rule -type of behavior, 
have been theoretically verified in numerous studies 
employing, e.g., SDFT.~\cite{hirose,reimann,aichinger}

\begin{figure}
\includegraphics[width=0.95\columnwidth]{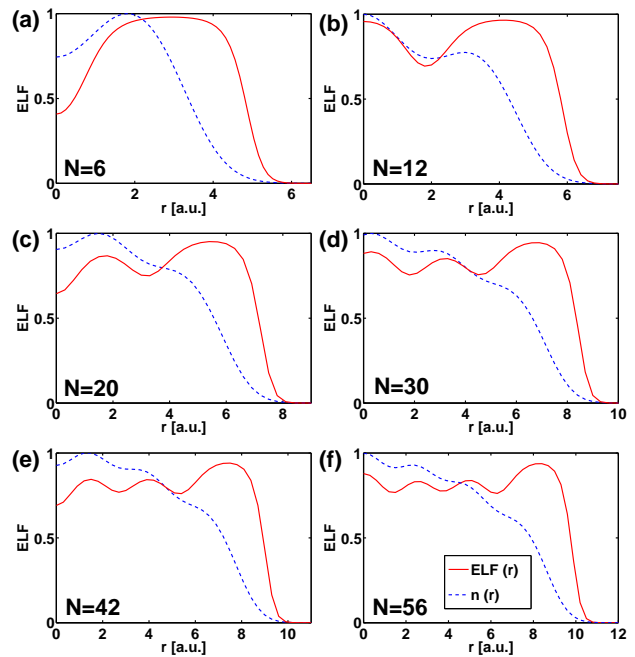}
\caption{(color online). ELF (red solid lines) with
the corresponding electron densities (blue dashed lines) for
two-dimensional closed-shell quantum dots. The maximum
density values are scaled to one.
}
\label{shells}
\end{figure}

Figure~\ref{shells}
shows the ELF (red solid lines) and the corresponding electron
densities (blue dashed lines) when $N=6,12,20,30,42,$ and $56$.
In QDs these ``magic'' electron numbers correspond to the first
closed-shell configurations when $N>2$. 
The shell structure is visualized clearly by the ELF.
We note, however, that the local maxima in the ELF do not
directly correspond to the filled energy shells. Instead, each 
peak can be associated with a doubly occupied single-electron 
state on the {\em highest} energy shell. This means that the 
probability of electron localization is highest close to 
the Fermi level. For example,
in a 12-electron QD shown in Fig.~\ref{shells}(b),
the highest energy shell has six electrons with 
$l=0,\pm 2$, leading to distinctive peaks in the ELF
at the center (corresponding to $l=0$) and at $r\sim 4$
(corresponding to $l=\pm 2$). Nevertheless, in this system
the number of energy shells is equal to
the number of  extrema in the ELF. As expected, 
the variation between the extrema decreases as a 
function of $N$, since the 
highest shell becomes relatively less dominant.
As seen in Fig.~\ref{shells}, the shell structure is visible
also in the electron density $n(r)$, but it is less clear
than in the ELF, especially at large electron numbers where
the structure is barely visible. 
The ELF instead provides an unambiguous visualization of the
shell structure, which is well plausible from its physical
nature discussed in Secs.~\ref{section:introduction} and
\ref{section:elf}.

\subsection{Quantum-dot molecules}
\label{subsection:qdm}

\begin{figure}
\includegraphics[width=0.95\columnwidth]{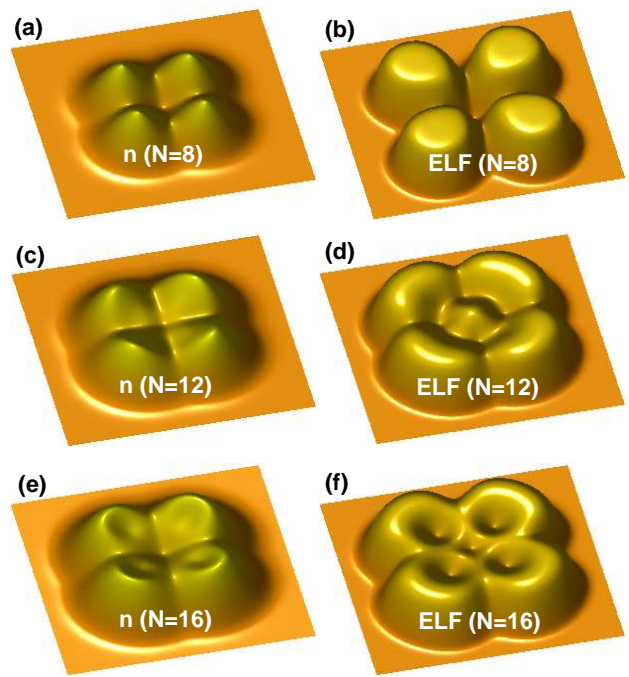}
\caption{(color online) Ground-state 
total densities and ELFs for four-minima 
quantum-dot molecules having $8$, $12$, and 
$16$ electrons, respectively. The total
spin $S=0$ in all the cases.
}
\label{qdm}
\end{figure}

The applicability of the ELF to visualize
pairs and bonds in molecules immediately
suggests to use the ELF for the 2D counterparts
commonly known as QD molecules (QDMs). 
Since the spin-qubit proposal of Loss and 
DeVinzenzo~\cite{loss} in 1998, coupled QD systems have 
attracted wide interest both 
experimentally~\cite{qdm_exp} 
and theoretically~\cite{qdm_theory}, particularly 
in terms of charge and spin manipulation.
Following the standard QDM definition, 
we write the external confining potential as
\begin{equation}
V_{\rm ext}({\mathbf r})=\frac{1}{2}\omega_0^2\,{\rm min}\left[\sum_j^M ({\mathbf r}-{\mathbf r}_j)^2\right],
\end{equation}
where $M$ is the number of potential minima
located at ${\mathbf r}_j=(x_j,y_j)$. 
The case of $M=1$, ${\mathbf r}_j=(0,0)$
is equal to a single harmonic QD considered 
above. Here we set $M=4$
and ${\mathbf r}_j=(\pm \sqrt{2},\pm \sqrt{2})$ with
$\omega_0=0.5$ a.u. This corresponds to
a square-symmetric QDM with four 
minima.~\cite{meripaper} 
Now, Fig.~\ref{qdm} 
shows the result of our SDFT calculations
for the total densities and ELFs when $N=8$ 
(a-b), $N=12$ (c-d), and $N=16$ (e-f), respectively. 
The ground-state total spin $S=0$ in all the cases 
shown. The densities of $N=8$ and $N=12$ QDMs look very
similar with four distinctive peaks, whereas the 
$N=16$ QDM shows more structure around the density maxima.
However, the ELFs of these QDMs are very different 
from each other, having four maxima with one (local)
minimum ($N=8$), five maxima with four minima ($N=12$), 
and four maxima with five minima ($N=16$). 
Despite the fact that the 
concept of chemical bonding seems less meaningful
in systems where the confining potential is fixed
(such as in 2D QD-systems),
the central ``basin'' (see Sec.~\ref{subsection:topology})
in Fig.~\ref{qdm}(d), for example, is very informative:
there is pronounced localization at the center, where the 
confining potential actually has a repulsive cusp.
Furthermore, this ground-state property is not at all 
visible in the bare electron density shown in 
Fig.~\ref{qdm}(c). Therefore, we find that the ELF 
reveals features in the electronic structure of
QDMs which are absent in the electron density -- a
fact also demonstrated in 3D.

\subsection{Maximum-density droplet}
\label{subsection:mdd}

\begin{figure}
\includegraphics[width=0.95\columnwidth]{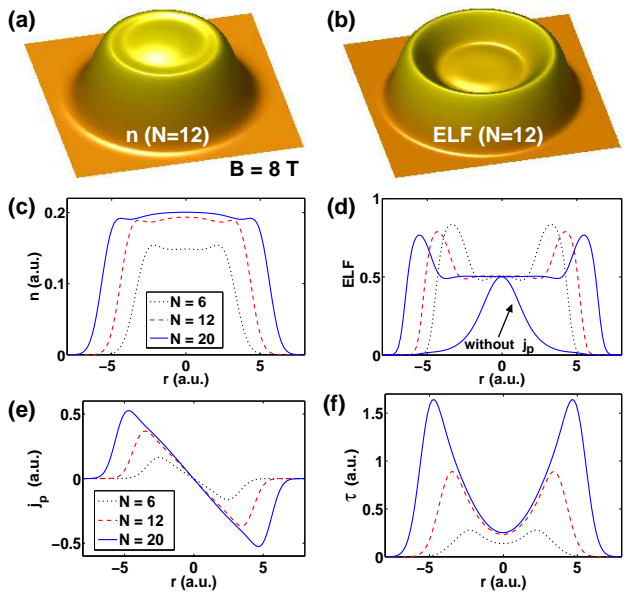}
\caption{(color online) (a,b) Electron density
and ELF of the maximum-density droplet
in a 12-electron quantum dot. (c,d) Intersections
of (a,b) together with the results for 
$N=6$ and $N=20$ quantum 
dots. (e,f) Paramagnetic current densities
and the kinetic-energy densities of the 
maximum-density-droplet states, respectively.
}
\label{mdd}
\end{figure}

Next we perform our analysis of the 2D ELF
for non-zero external magnetic fields ($B>0$).
As the first example we consider a
maximum-density droplet (MDD)~\cite{macdonald,oosterkamp}
of a single parabolic QD. The MDD is a fully-polarized
($S=N/2$) state, where the electrons
have consecutive angular momenta from
$l=0$ to $l=-N+1$. Each electron
is accompanied by a so-called {\em Pauli vortex}
which corresponds to a change of $2\pi$ in 
the phase of the many-electron wave function.  
Hence, the MDD state is a unique finite-size
counterpart of the filling-factor $\nu=1$ 
quantum Hall state of the uniform 2D 
electron gas. Figures~\ref{mdd}(a)
and (b) show the electron density and the ELF
of the MDD state in a 12-electron parabolic 
QD at $B=8$ T. The MDD density has the well-known
flat shape,~\cite{macdonald,oosterkamp} whereas the ELF
is characterized by a flat interior and 
localization around the edge of the QD. 
Intersections of the density and the ELF
are plotted in Figs.~\ref{mdd}(c-d) (red dashed lines), 
together with the results for 6 and 20-electron QDs
at $B=6$ T and $8$ T, respectively.
Interestingly, the values of the ELF are
very close to $1/2$ in the interior of the MDDs.
Thus, from Eq.~(\ref{eq:ELF}) we find 
$C_{\sigma}({\bf r})\approx C_{\sigma}^{\rm HEG}({\bf r})$
in this regime, i.e., the kinetic-energy density
of the uniform 2D electron gas. The result suggests that
the localization of electrons and vortices 
compensate each other, which is well
plausible considering the $\nu=1$ character
of the MDD state. We may thus expect that at
higher magnetic fields, corresponding to the 
fractional quantum-Hall regime ($\nu<1$),
ELF values below $1/2$ can be found at positions
of high vorticity.

Figure~\ref{mdd}(d) shows also the ELF
of the $N=20$ QD calculated without
the current-density term $j_p$ in Eq.~(\ref{eq:csigma1det}).
That curve equals to the full ELF at $r=0$
where the current density is zero, but decays
exponentially at larger $r$. Obviously, 
that result does not capture the correct behavior
of the MDD state. The dramatic difference 
demonstrates the importance of the current-density 
dependence in the ELF expression, already
at relatively moderate magnetic fields.
In Fig.~\ref{mdd}(e) we plot $j_p$ which
increases linearly due to the successive
increase in the angular momenta of the
KS states in the MDD (see above).

\subsection{Vortex structures}
\label{subsection:vortex}

\begin{figure}
\includegraphics[width=0.95\columnwidth]{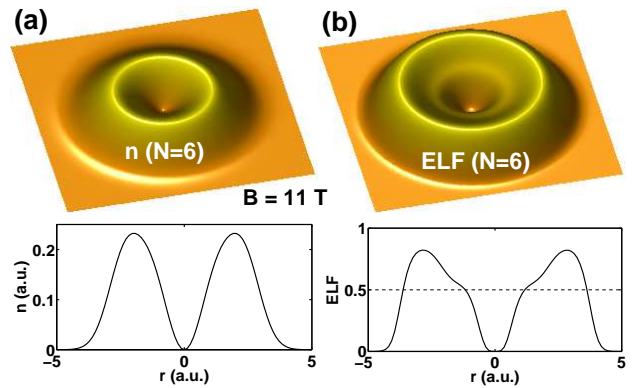}
\caption{(color online). Electron density (a) and the ELF (b)
of a single-vortex solution in a six-electron parabolic 
quantum dot at $B=11$ T. The lower panel shows middle intersections 
of the figures in the upper panel.}
\label{singlevortex}
\end{figure}

Increasing the magnetic field above
the MDD limit at $\nu=1$, which, as seen above,
corresponds to ELF=$1/2$, leads to
localization of vortices. Depending on 
the QD geometry, the vortices may form
clusters~\cite{vortex_henri,vortex_manninen,vortex_recta,vortex_peeters} 
or merge together yielding multiple 
phase quantization.~\cite{vortex_giant}
In Fig.~\ref{singlevortex}
we show the electron density and the ELF
of a single-vortex solution in a six-electron 
parabolic QD at $B=11$ T. In this case the
vortex is localized at the center and directly 
visible as a hole in the density. As expected, 
the ELF shows a similar strucure. The ELF
has a bump with a value $\sim 1/2$, located between 
the center and the edge of the QD. This region can
be interpreted to have, on the average, a local balance 
between electrons and (Pauli-)vortices, and it
separates the localized vortex (ELF$=0$) from the
edge having high electron localization (ELF$\approx 0.8$).

\begin{figure}
\includegraphics[width=0.95\columnwidth]{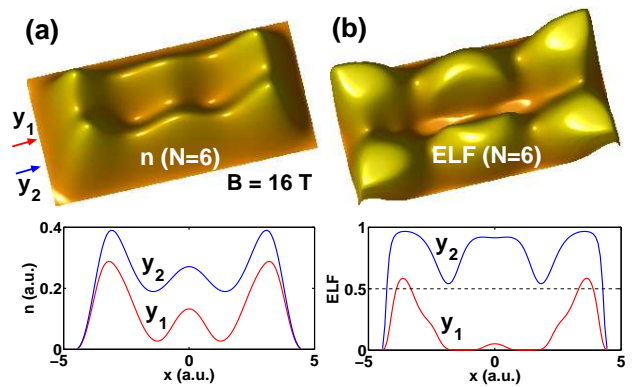}
\caption{(color online). Electron density (a) and the ELF (b)
of a two-vortex solution in a six-electron rectangular
quantum dot at $B=16$ T. The lower panel shows intersections
of the upper figures at $y_1$ and $y_2$, marked in the upper 
right figure.}
\label{rectavortex}
\end{figure}

Increasing the magnetic field further leads 
to the formation of more vortices. Figure~\ref{rectavortex}
shows a two-vortex structure in a six-electron 
rectangular (hard-wall) QD at $B=16$ T. The system
has been studied in detail in Refs.~[\onlinecite{vortex_recta}] and
[\onlinecite{recta2}]. Again, the vortices are 
seen as zeros in the ELF, whereas the density is {\em not}
exactly zero at these positions. 
Actually, 
the numerically exact density at the vortices 
is even further above zero than 
the SDFT density due to the 
multiconfigurational nature of the many-electron
wave function.~\cite{vortex_recta}
Nevertheless, the ELF=$0$ result at the
vortex positions is plausible, reflecting 
again the difference between the probable
{\em location} (density) and the {\em localization} 
(ELF) of the electrons as discussed in 
Sec.~\ref{section:introduction}.
Along the edge of the QD, the ELF in
Fig.~\ref{rectavortex}(b) shows six
clear peaks with maximum values close to one, 
separated by local minima where ELF$\sim 1/2$.
Thus, in this QD the ELF reveals the Wigner
crystallization, i.e., localization of electrons
around their classical positions which are
determined by the geometry of the system. 
This effect is considerably less pronounced
in the electron density shown in 
Fig.~\ref{rectavortex}(a).



\section{Concluding remarks}
\label{sec:summary}

The electronic structure of a many-electron system is fully
characterized by its many-body wave function. In order to acquire an
intuitive visual understanding of the system, however, we must look at
simpler objects -- integrated magnitudes such as the electronic
density, which lives in a lower dimensional space. Unfortunately, the
density does not fully reveal all the intricacies of the electronic
structure, even if we know that it contains all the information.

In the past, the ELF has proven to be a useful companion to the
density in the task of providing us with insightful intuition on the
electronic structure of molecules. In this work we have defined the
ELF in 2D, and we have demonstrated that its visualization, 
in addition to that of the density, helps to understand
the electronic structure of 2D systems such as semiconductor
quantum dots. We have shown the usefulness
of the ELF to visualize the shell structure, as well as 
the bond-like features in coupled systems.
In particular, we have found that in magnetic fields the ELF
can be used as a measure of vorticity, revealing the
local relation between the localization of electrons and 
vortices in the system. In this context, we have shown 
that the current-dependent term, which has been neglected
in previous 3D studies, has a major contribution to the ELF 
in magnetic fields. We expect that alongside the rapid
technological developments in the fabrication and manipulation
of low-dimensional systems, the ELF will prove to be
a universally applicable theoretical tool to obtain detailed information of 
both static and dynamic many-particle properties.

\begin{acknowledgments}
This work was supported by the EU's Sixth Framework
Programme through the Nanoquanta Network of Excellence 
(NMP4-CT-2004-500198), Barcelona Mare Nostrum Center, the Academy of
Finland, the Finnish Academy of Science and Letters through the Viljo,
Yrj{\"{o}}, and Kalle V{\"{a}}is{\"{a}}l{\"{a}} Foundation, and the
Deutsche Forschungsgemeinschaft through SFB 658. 
\end{acknowledgments}

\end{document}